# NEUTRON STARS WITH A QUARK CORE. I. EQUATIONS OF STATE

G. B. Alaverdyan[1], A. R. Harutyunyan[2], and Yu. L. Vartanyan[3]

*An extensive set of realistic equations of state for superdense matter with a quark phase transition is derived on the basis of the three equations of state for neutron matter and the eight variants of strange quark-gluon plasmas in the MIT quark bag model. The characteristics of the phase transitions are described and the calculated equations of state with a density jump are studied in detail.*

## 1. Introduction

At superhigh densities a phase transition can occur from a state in which the quarks contained within baryons to a free quark plasma state. A phase transition to so-called strange quark matter in a superdense nuclear plasma has been considered [1].

Because of difficulties with the theory of strong interactions, the quark phase is described using models. The absence of an exact form for the equations of state in both the neutron and the quark-gluon regions makes it important to compare the dependences of the parameters of the corresponding superdense heavenly objects on the equations of state that are used.

Computational models of strange stars have been reported in a number of papers and they have been analyzed from various standpoints [2-7]. Far fewer papers, however, have been devoted to studies of stellar configurations with a density jump [8-12].

In this regard, we should take note of the most complete calculations [13-16] of models with a "mixed phase" that contains quark formations with various configurations in the form of droplet-, rod-, and plate-shaped structures and which assume a continuous variation in the pressure and density within the region where the quark phase exists [17]. These papers have shown that the formation of a mixed phase of quark and nuclear matter may turn out to be more or less preferable in terms of energy than an ordinary first order phase transition from nucleonic to quark matter, depending on the magnitudes of the local surface and Coulomb energies associated with the formation of the quark and nuclear

———————————
*Yerevan State University*

E-mail: 1) galaverdyan@ysu.am
       2) anharutr@ysu.am
       3) yuvartanyan@ysu.am

structures in the mixed phase [13,14,16]. Because of the uncertainty in the value of the surface tension of strange quark matter, it is not possible at this time to establish unambiguously which of the above variants will be actually realized. In the following we examine a case which assumes a surface tension that yields a first order phase transition with the possible coexistence of two phases.

In this paper we present a detailed study of layered configurations with a density jump that contain a strange quark core. By combining the three equations of state of neutron matter with different variants of the strange quark-gluon plasma calculated in terms of the MIT quark bag model [18], we have constructed an extensive set of realistic equations of state with a quark phase transition.

The calculated equations of state are used to integrate the system of Tolman-Oppenheimer-Volkov equations and find such parameters of neutron stars with strange cores as the mass, radius, relativistic moment of inertia, and red shift from the surface of a star, as well as the mass and radius of the quark core within the allowed range of central pressures. These results will be published separately (in the second part of this paper).

## 2. Equations of state

We have examined the transition to the quark phase in nuclear plasmas using a quark "bag" model for the strange quark matter. Here, depending on the inadequately known strong interaction parameters, the energy per baryon may have either a negative or positive minimum. Whereas in the first case the strange quark matter cannot exist in a state of thermodynamic equilibrium with a baryon component (the case where strange stars are formed), in the second, a thermodynamic equilibrium is possible between the quark and nucleonic components (as occurs in neutron stars with a quark core). We have studied an extensive set of suitable equations of state for different variants of the neutron and quark phases. In particular, for equations of state for the nucleonic component we considered models of neutron matter obtained in the framework of a relativistic nuclear field [19], on the basis of potentials [20-22] which include the effect of two-particle correlations in the so-called $\lambda^{00}$ approximation [23-24]. These equations of state, which are used in the near-nuclear and supernuclear density ranges ($3.56 \times 10^{13} < \rho < 4.81 \times 10^{14}$ g/cm$^3$) and denoted by "Bonn" and "HEA" in this paper, are matched with the equation of state [25] for superdense matter at the lower densities. In the calculations we have also used the BJ-V potential [26] as a more "rigorous" equation of state for neutron matter than the "Bonn" and "HEA" forms.

By combining these three equations of state for the neutron component with eight variants of the strange quark-gluon plasma calculated for a realistic range of quark bag parameters ($B \approx 55 \div 60$ MeV/fm$^3$, $m_s \approx 175 \div 200$ MeV, $\alpha_c \approx 0.5 \div 0.6$), we have obtained a set of equations of state with a quark phase transition.

Table 1 lists the notations for the different variants of the quark component, as well as the value of the energy $\varepsilon_{min}$ per baryon and the baryon density $n_{min}$ at the minimum point. (For details on the description of strange quark matter see Ref. 27.) Here $B$ is the "bag" constant that characterizes the vacuum pressure and ensures confinement, $\alpha_c$ is the quark-gluon interaction constant, and $m_s$ is the mass of the strange quark.

Table 1 shows that the variation in $n_{min}$ is less than 7% over all the variants. The depth $\varepsilon_{min}$ of the energy well is more sensitive to the parameters of the quark model. A change in one of the parameters by ten percent leads to a change in $\varepsilon_{min}$ by more than a factor of two.

Whereas the average energy e per baryon increases monotonically with the baryon concentration n in the nucleonic phase, in the quark phase this same quantity has a minimum $e_{min}$ corresponding to quark creation. Figure 1 shows the average energy per baryon as a function of the specific volume $1/n$ for the different quark plasma models. (The letters next to the curves correspond to the notation of Table 1.) Also shown in this figure are plots for the three variants of the equation of state for neutron matter (1 corresponds to "HEA," 2 to "Bonn," and 3 to "BJ-V").

TABLE 1. Variants of the Quark-Gluon Plasma Used in this Paper

| Notation | $m_s$, MeV | $B$, MeV/fm$^3$ | $\alpha_c$ | $\varepsilon_{min}$, MeV | $n_{min}$, fm$^{-3}$ |
|---|---|---|---|---|---|
| a | 175 | 55 | 0.5 | 10.44 | 0.258 |
| b | 200 | 55 | 0.5 | 20.71 | 0.263 |
| c | 175 | 55 | 0.6 | 28.61 | 0.258 |
| d | 175 | 60 | 0.5 | 28.97 | 0.276 |
| e | 200 | 55 | 0.6 | 38.24 | 0.259 |
| f | 200 | 60 | 0.5 | 39.12 | 0.282 |
| g | 175 | 60 | 0.6 | 47.44 | 0.275 |
| h | 200 | 60 | 0.6 | 56.90 | 0.277 |

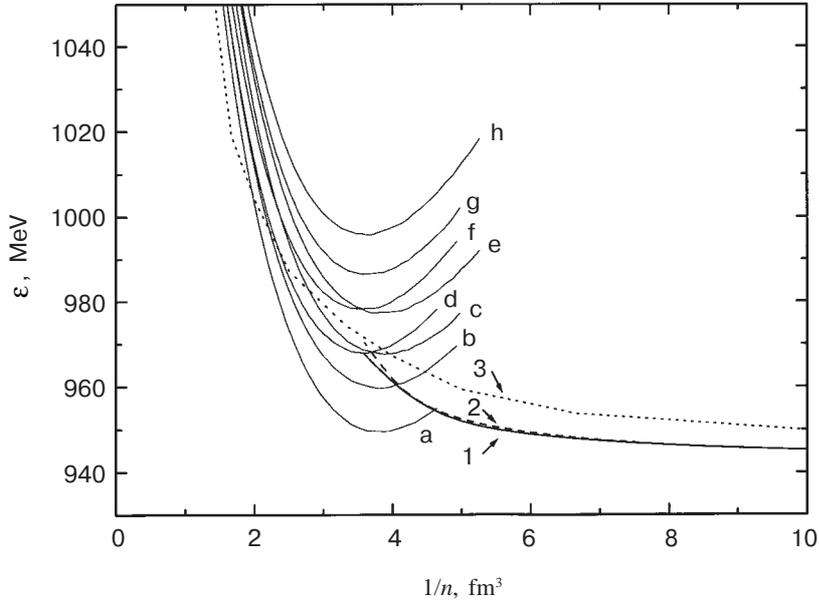

Fig. 1. The average energy per baryon as a function of the specific volume $1/n$ for neutron matter and a quark-gluon plasma. Curves 1, 2, and 3 represent the "HEA," "Bonn," and "BJ-V" equations of state, respectively. The curves labelled with letters represent the equations of state for quark matter listed in Table 1.

An analysis showed that the conditions for realizing a phase transition to the quark phase and thermodynamic equilibrium with the nucleonic component are possible only for eleven of the twenty four equations of state considered, and for the three models of neutron matter discussed above, equilibrium and simultaneous coexistence with the quark variants *e, g,* and *h*, for which $\varepsilon_{min}$ is high, are impossible. We now clarify this point. It is known that a first order phase transition from the nucleonic to the quark phase takes place at pressure $P_0$ if the Gibbs conditions are satisfied: $\mu_N(P_0) = \mu_Q(P_0)$ and $\mu_N(P) > \mu_Q(P)$ for $P > P_0$, where $P_0$ is the pressure at the phase transition point, and $\mu_N$ and $\mu_Q$ are the chemical potentials of the nucleonic and quark phases, respectively. Figure 2 shows the chemical potentials as functions of pressure for nucleonic matter with the Bonn equation of state (smooth curve) and different quark models. The dashed curves represent $\mu_Q(P)$ for the quark models. The models are indicated in accordance with the notation of Table 1. The corresponding curves are similar for the other two nucleonic equations of state. The intersection of the dashed curves with the smooth curve corresponds to a phase transition from the nucleonic component to the corresponding quark model. Figure 2 shows that the quark models with high $\varepsilon_{min}$ (models *e, g,* and *h*) do not have phase transition points. For these models the situation is analogous to that of strange stars with the distinguishing feature that in the case of strange stars, $\varepsilon_{min} < 0$, whereas here $\varepsilon_{min} > 0$.

Values of the pressure $P_0$, energy densities $\rho_N$ and $\rho_Q$, and baryon densities $n_N$ and $n_Q$ in the nucleonic and quark phases, respectively, at the phase transition point are given in Table 2. The last column lists the density jump parameter $\lambda$, which is given by $\lambda = \rho_Q / (\rho_N + P_0/c^2)$ for relativistic models [28]. Its value determines the stability regions on plots of star mass $M$ as a function of the central pressure $P_c$ for layered models.

The eleven equations of state with phase transitions have been assigned notations in which a number refers to the nucleonic component (1 corresponds to HEA, 2 to Bonn, and 3 to BJ-V) and a letter, to the quark model whose parameters are given in Table 1. Thus, for example, the model labelled 2d is the equation of state model for which the nucleonic component is the Bonn variant and the quark model is variant d of Table 1.

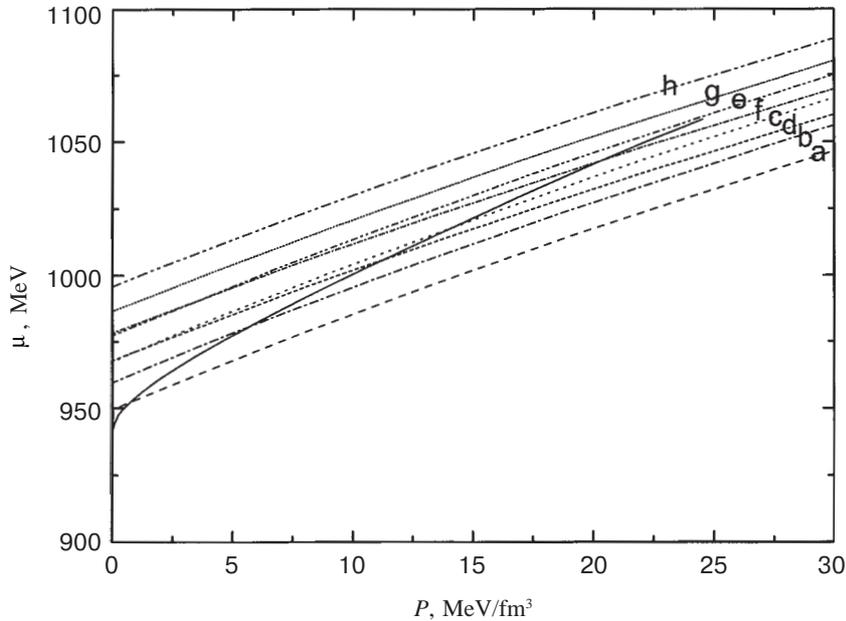

Fig. 2. Chemical potentials as a function of pressure for neutron matter with the Bonn equation of state (smooth curve) and different variants of the quark matter (per the notation of Table 1).

TABLE 2. Characteristic Phase Transition Points for Equations of State with a Density Jump

| Equation of state | $P_0$, MeV/fm$^3$ | $\rho_N$, $10^{14}$ g/cm$^3$ | $\rho_Q$, $10^{14}$ g/cm$^3$ | $n_N$, fm$^{-3}$ | $n_Q$, fm$^{-3}$ | $\lambda$ |
|---|---|---|---|---|---|---|
| 3a | 0.199 | 0.861 | 4.421 | 0.051 | 0.262 | 5.11 |
| 2a | 0.758 | 2.023 | 4.454 | 0.120 | 0.264 | 2.19 |
| 1a | 0.762 | 2.073 | 4.454 | 0.123 | 0.264 | 2.13 |
| 3b | 0.796 | 1.725 | 4.531 | 0.102 | 0.266 | 2.60 |
| 3d | 5.291 | 3.859 | 5.103 | 0.225 | 0.296 | 1.291 |
| 2b | 5.674 | 3.513 | 4.827 | 0.207 | 0.283 | 1.34 |
| 1b | 5.979 | 3.478 | 4.845 | 0.205 | 0.284 | 1.35 |
| 3c | 6.013 | 4.069 | 4.777 | 0.237 | 0.277 | 1.14 |
| 2d | 11.219 | 3.958 | 5.449 | 0.232 | 0.316 | 1.31 |
| 2c | 14.107 | 4.120 | 5.249 | 0.241 | 0.304 | 1.20 |
| 2f | 20.273 | 4.473 | 6.086 | 0.261 | 0.348 | 1.26 |

Table 2 shows that a phase transition to a quark state is possible for all three of the neutron equations of state considered here only in the case of quark models a and b, which have the lowest values of $\varepsilon_{min}$. With the next two quark models (c and d) a transition is only possible for the Bonn and BJ-V equations of state (variants 2 and 3), while for quark model f, with a relatively high $\varepsilon_{min}$, it can occur only for the Bonn variant.

As Table 2 implies, the parameter $\lambda$ for quark model a exceeds 1.5 for all three nucleonic equations of state, and for model b, only in the case of the BJ-V nucleon model. An instability region shows up in a plot of the mass as a function of the central pressure, $M(P_c)$, for these four models owing to a transition to a quark phase (the region of neutron stars with low-mass quark cores).

For the remaining seven equation of state models, which correspond to $\lambda < 1.5$, a transition to a quark phase only leads to a break in the $M(P_c)$ curve without a change in the sign of the derivative, and no instability develops.

It can be seen from Table 2 that for the phase transitions with $\lambda > 1.5$, $n_Q$ only exceeds $n_{min}$ slightly (See Table 1.), while $n_N$ is lower than the baryon concentration at the nuclear density $n_0$ ($n_0 = 0.14$ fm$^{-3}$). In the case of the quark models with relatively high $\varepsilon_{min}$ (for which $\lambda < 1.5$), on the other hand, $n_N$ is substantially higher than $n_0$. For these models with their relatively low density jumps, the pressure $P_0$ at which the phase transition occurs turns out to be extremely high.

Figure 3 contains plots of the equations of state with density jumps in $(P, \rho)$ space for all the variants considered here. These equations of state yields models of neutron stars with strange quark cores.

This work was supported by the Armenian National Science and Education Foundation (ANSEF Grant No. PS 140) and undertaken in the framework of topic No. 0842 with the financial support of the Ministry of Education of the Republic of Armenia.

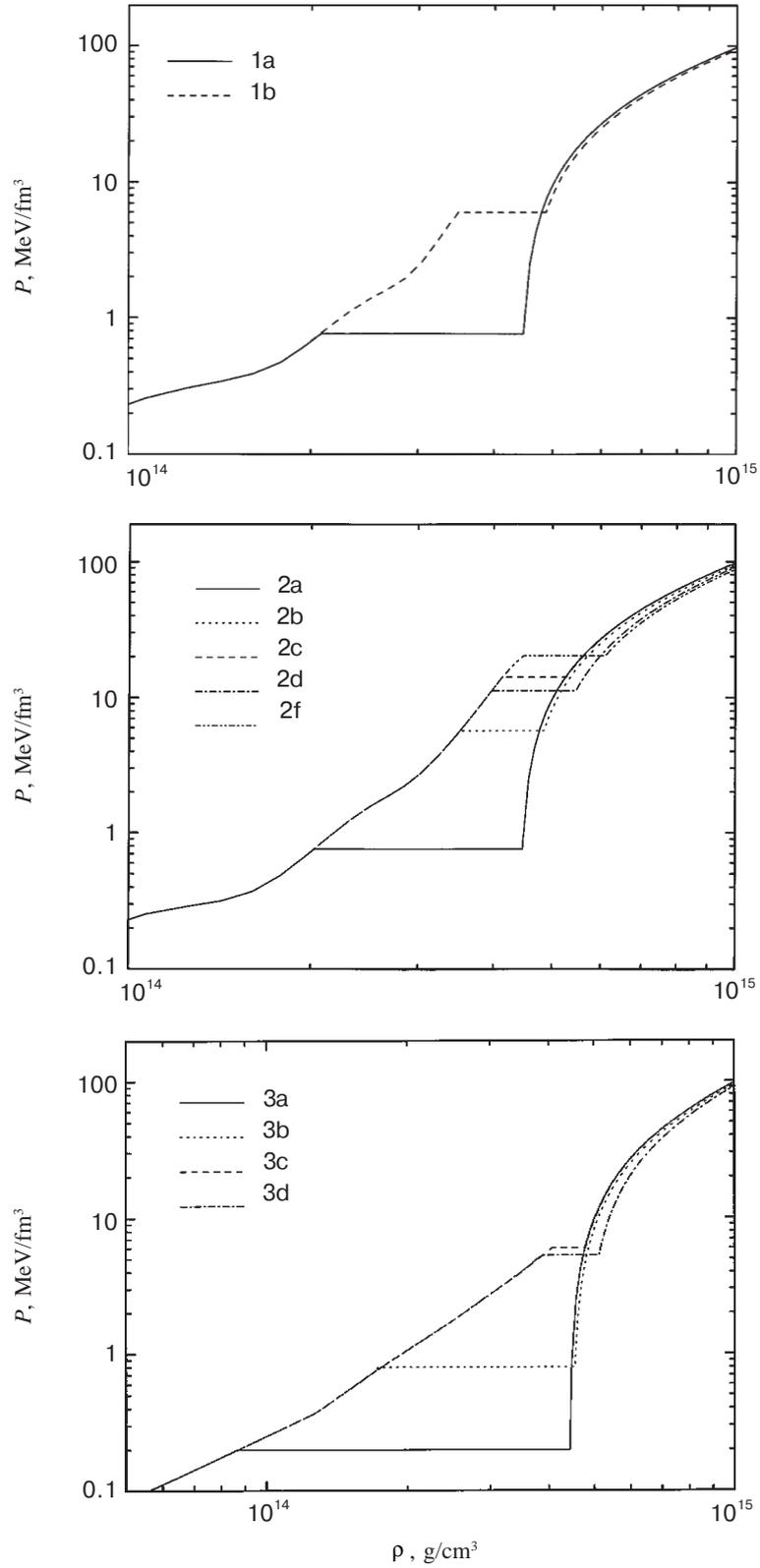

Fig. 3. Equations of state with a first order quark phase transition. The notation is the same as in Figs. 1 and 2 (HEA (1), Bonn (2), BJ-V (3) and the letters refer to the quark models).

**REFERENCES**

1. E. Witten, *Phys. Rev.* **D30**, 272 (1984).
2. C. Alcock, E. Farhi, and A. Olinto, *Astrophys. J.* **310**, 261 (1986).
3. P. Haensel, J. L. Zdunik, and R. Schaeffer, *Astron. Astrophys.* **160**, 121 (1986).
4. L. A. Kondratyuk, M. I. Krivoruchenko, and B. V. Martem'yanov, *Pis'ma v Astron. Zh.* **16**, 954 (1990).
5. F. Weber and N. K. Glendenning, LBL-33066 (1992).
6. Yu. L. Vartanyan, A. R. Arutyunyan, and A. K. Grigoryan, *Pis'ma v Astron. Zh* **21**, 136 (1995).
7. S. B. Khadkikar, A. Mishra, and H. Mishra, *Mod. Phys. Lett.* **A10**, 2651 (1995).
8. P. A. Carinhas, *Astrophys. J.* **412**, 213 (1993).
9. G. B. Alaverdyan, A. R. Arutyunyan, Yu. L. Vartanyan, and A. K. Grigoryan, Dokl. AN RA **95**, 98 (1995).
10. G. B. Alaverdyan, A. R. Arutyunyan, and Yu. L. Vartanyan, *Astrofizika* **44**, 323 (2001).
11. G. B. Alaverdyan, A. R. Arutyunyan, and Yu. L. Vartanyan, *Pis'ma v Astron. Zh.* **28**, 29 (2002).
12. A. R. Arutyunyan, *Astrofizika* **45**, 307 (2002).
13. H. Heiselberg, C. J. Pethick, and E. F. Staubo, *Phys. Rev. Lett.* **70**, 1355 (1993).
14. C. P. Lorenz, D. G. Ravenhall, and C. J. Pethick, *Phys. Rev. Lett.* **70**, 379 (1993).
15. N. K. Glendenning, astro-ph/9706236 (1997).
16. H. Heiselberg and M. Hjorth-Jensen, nucl-th/9902033 (1999).
17. N. K. Glendenning, *Phys. Rev.* **D46**, 1274 (1992).
18. A. Chodos, R. L. Jaffe, K. Johnson, C. B. Thorn, and V. F. Weisskopf, *Phys. Rev.* **D9**, 3471 (1974).
19. F. Weber, N. K. Glendenning, and M. K. Weigel, *Astrophys. J.* **373**, 579 (1991).
20. R. Machleidt, K. Holinde, and Ch. Elster, *Phys. Rep.* **149**, 1 (1987).
21. K. Holinde, K. Erkelenz, and R. Alzetta, *Nucl. Phys.* **A194**, 161 (1972).
22. K. Holinde, K. Erkelenz, and R. Alzetta, *Nucl. Phys.* **A198**, 598 (1972).
23. P. Poschenrieder and M. K. Weigel, *Phys. Lett.* **B200**, 231 (1988).
24. P. Poschenrieder and M. K. Weigel, *Phys. Rev.* **C38**, 471 (1988).
25. G. S. Bisnovatyi-Kogan, *Physical Problems in the Theory of Stellar Evolution* [in Russian], Nauka, Moscow (1989).
26. R. C. Malone, M. B. Johnson, and H. A. Bethe, *Astrophys. J.* **199**, 741 (1975).
27. E. Farhi and R. L. Jaffe, *Phys. Rev.* **D30**, 2379 (1984).
28. M. J. Lighthill, *Mon. Notic. Roy. Astron. Soc.* **110**, 339 (1950).